# EFFECT OF INTER PACKET DELAY IN PERFORMANCE ANALYSIS OF COEXISTENCE HETEROGENEOUS WIRELESS PACKET NETWORKS

G.M.Tamilselvan[1] and Dr.A.Shanmugam[2]

[1] Department of Electronics and Communication Engineering, Bannariamman Institute of Technology, Sathyamangalam, Tamilnadu, India
`tamiltamil@rediffmail.com`

[2] Principal, Bannariamman Institute of Technology, Sathyamangalam, Tamilnadu, India
`dras@yahoo.com`

## ABSTRACT

*As the explosive growth of the ISM band usage continues, there are many scenarios where different systems operate in the same place at the same time. One of growing concerns is the coexistence of heterogeneous wireless network systems. For the successful deployment of mission-critical systems such as wireless sensor networks, it is required to provide a solution for the coexistence. In this paper, we propose a new scheme using inter packet delay for the coexistence of IEEE 802.15.4 LRWPAN and IEEE 802.11b WLAN. To evaluate the effectiveness of the proposed scheme, measurement and simulation study are conducted using Qualnet 4.5 simulation software. The simulation results show that the proposed scheme is effective in performance improvement for coexistence network of IEEE 802.15.4 for various topologies.*
.

## KEYWORDS

*Coexistence, Heterogeneous wireless network, IEEE 802.15.4, IEEE 802.11b, Inter packet delay.*

## 1. INTRODUCTION

The Industrial, Scientific and Medical (ISM) band is widely used among popular wireless network standards such as IEEE 802.15.4 Low-Rate Wireless Personal Area Network (LRWPAN), IEEE 802.11b Wireless Local Area Network (WLAN), IEEE 802.15.3, and Bluetooth. Because of the mobility and ubiquitous deployment of wireless systems, there are many scenarios where different systems operate in the same place at the same time. Hand-held PDA can use a Bluetooth device to connect to a laptop with 802.11b WLAN. The ISM band is also used by home appliances such as microwave ovens. The microwave oven in the house can be turned on when cordless phone is being used.

Coexistence is defined as "the ability of one system to perform a task in a given shared environment where other systems may or may not be using the same set of rules" .Especially, for mission-critical applications using wireless systems; the coexistence becomes a top priority issue in system design. For example, if 802.15.4 sensor network system is to be deployed in the hospital building for emergency medical care, a main design issue will be providing the coexistence of 802.15.4 and other wireless systems. In case the other system causes radio channel interference, the sensor network system can not continue the normal operation and may lose critical information such as emergency patient vital signals and emergency patient information.

Growing concern is in the coexistence of 802.1.5.4 and 802.11b. There are many practical situations and scenarios where 802.15.4 and 802.11b systems operate simultaneously. An interesting system is sensor network system employing IEEE 802.15.4(WPAN) technology. Recently, Wireless personal area networks (WPAN) are one of most essential technologies for implementing ubiquitous computing. Wireless personal area networks are used to convey information over relatively short distances. Unlike wireless local area networks, connections effected via WPANs involve little or no infrastructure. This feature allows small, power-efficient, inexpensive solutions to be implemented for a wide range of devices.

The IEEE 802.15.4 would be widely adopted for various applications such as detection, remote control, tracking, and monitoring [1] [2]. The scope of IEEE 802.15.4 [3][4] is to define the physical layer (PHY) and medium access control (MAC) sub layer specifications for low data rate wireless connectivity with fixed, portable, and moving devices with no battery or very limited battery consumption requirements typically operating in the personal operating space (POS) of 10 m. It is foreseen that, depending on the application, a longer range at a lower data rate may be an acceptable trade-off. The purpose of IEEE 802.15.4 is to provide a standard for ultra-low complexity, ultra-low cost, ultra-low power consumption, and low data rate wireless connectivity among inexpensive devices.

Many practical wireless sensor network systems cover a large area of interest. The examples include wild life habitat monitoring, hospital emergency medial care and health monitoring, forest fire detection and tracking, traffic monitoring and others. Because those systems have a large coverage area and the same area can have other human activities such as residences and leisure activities, it is reasonably assumed that the 802.15.4 system will be operating with other systems. It is interesting to note that the effect caused by radio interference is not reciprocal when multiple wireless systems operate simultaneously. It is because of the difference in radio transmission range. 802.11b uses a longer range radio than 802.15.4 system. 802.11b WLAN has radio range of 100 m and 802.15.4 LR-PAN has radio range of 10 m [12]. Thus, 802.11b can give radio interference to 802.15.4 system in a large area and from a long distance. Therefore, large-scale 802.15.4 based sensor network system is vulnerable to the interference from 802.11b. Moreover, 802.11b systems are employed in many portable devices including hand-held Personal Data Assistant (PDA) and laptop computers. Due to the omnipresence and mobility of those systems, there is a high chance of operating 802.15.4 and 802.11b in the same environment. There are many situations, where 802.15.4 and 802.11b need operate in the same system. For example, 802.15.4 wireless sensor nodes forward the sensing data to a laptop, which will be send the collected data over 802.11b WLAN to the central computer for processing and further analysis.

In the coexistence of IEEE 802.15.4 and IEEE 802.11b, the main concern is the performance degradation of IEEE 802.1.5.4 caused by the interference of IEEE 802.11b. A measurement study reported that over 92 % of the 802.15.4 frames were lost by the interference of IEEE 802.11b [13]. In this paper, we propose a scheme using inter packet delay to solve the performance degradation of IEEE 802.15.4. Especially, the proposed scheme is intended to support coexistence performance issue for IEEE 802.15.4 multi-hop network.

The rest of this paper is organized as follows: Section 2 summarizes the related works. The proposed scheme is presented in Section 3. Simulation results are discussed in Section 4. Finally, we conclude our paper.

## 2. RELATED WORK

Figure 1 shows the operational frequency spectrum of both IEEE 802.15.4(ZigBee) and IEEE 802.11b (WLAN) networks. A WLAN system has eleven channels. Each channel occupies 22

MHz and up to 3 separate channels can be simultaneously used without any mutual interference. Channels 1, 6, and 11 can be used for neighboring IEEE 802.11 WLAN Access Points (APs), as shown in Figure 1, to mitigate the interference. On the other hand, ZigBee networks have sixteen channels in 2.4 GHz band which can be used simultaneously without any mutual interference among them. Since the transmission power of WLAN is usually 100 times larger than that of ZigBee networks, we focus on the effect of interference from WLAN to ZigBee.

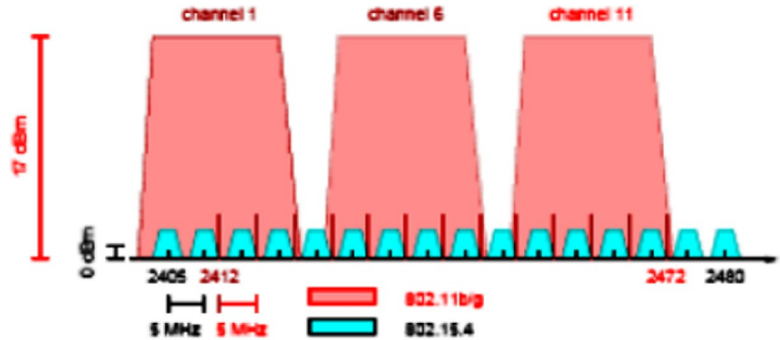

Figure.1. Frequency Spectrum of ZigBee and WLAN Networks

In IEEE 802.15.4 standard, a transmission between PAN coordinator and devices is performed inside the 2.4 GHz ISM band, at 250 kbps, and exploiting one of the 16 available channels. As shown in table 1, such channels have a 3 MHz bandwidth and are uniformly distributed within the ISM band.

Table 1. 2.4GHz ISM Band, IEEE 802.15.4 and IEEE 802.11 Channels

|  | IEEE 802.11b | | IEEE 802.15.4 | | IEEE 802.11b | | IEEE 802.15.4 | |
|---|---|---|---|---|---|---|---|---|
|  | Ch. | Freq. (GHz) | Ch | Freq. (GHz) | Ch | Freq. (GHz) | Ch | Freq. (GHz) |
| 2.4 GHz ISM Band | 1 | 2.401- 2.423 | 1 | 2.405 | 9 | 2.441-2.463 | 9 | 2445 |
|  | 2 | 2.404- 2.426 | 2 | 2.410 | 10 | 2.446-2.468 | 10 | 2450 |
|  | 3 | 2.411-2.433 | 3 | 2.415 | 11 | 2.451-2.473 | 11 | 2455 |
|  | 4 | 2.416-2.438 | 4 | 2.420 | 12 |  | 12 | 2460 |
|  | 5 | 2.421-2.443 | 5 | 2.425 | 13 |  | 13 | 2465 |
|  | 6 | 2.426-2.428 | 6 | 2.430 | 14 |  | 14 | 2470 |
|  | 7 | 2.431-2.453 | 7 | 2.435 |  |  | 15 | 2475 |
|  | 8 | 2.436-2.458 | 8 | 2.440 |  |  | 16 | 2480 |

Some related researches study the coexistence problem between the IEEE 802.15.4 and the 802.11b [5],[6].In [5], the packet error rate (PER) of the IEEE 802.15.4 under the IEEE 802.11b and IEEE 802.15.1 is obtained by experiments only. In [6], the impact of an IEEE 802.15.4 network on the IEEE 802.11b devices is analyzed. Channel Conflict Probabilities between IEEE 802.15 based Wireless Personal Area Networks is modelled in [7]. Packet Error Rate of IEEE 802.15.4 under IEEE 802.11b interference is analyzed in [8].In [9] Packet Error Rate of IEEE 802.11b under IEEE 802.15.4 interference is analyzed. In [10] channel conflict probabilities between IEEE 802.11b and IEEE 802.15.4 have been modelled. In [11] channel collision between IEEE 802.15.4 and IEEE 802.11b for circular and grid topology is analysed with the mobility model. To the best knowledge of the authors, performance analysis of coexistence heterogeneous network for circular, grid and random topology by varying the inter packet delay has not been discussed in the literature.

## 3. PROPOSED SCHEME

In this paper, we propose an inter packet delay based analysis for the performance metrics such as data received with errors, throughput, average End-End delay and average jitter of IEEE 802.15.4. We consider a heterogeneous network with circular, grid and random topology.

Here the performance of IEEE 802.15.4 under the interference of IEEE 802.11b is analyzed using Qualnet 4.5 simulation. For simulation, the slotted CSMA/CA of the IEEE 802.15.4 model is developed using Qualnet 4.5.The scenario of coexistence heterogeneous network for circular, grid and random topology is shown in figure 2(a-c).

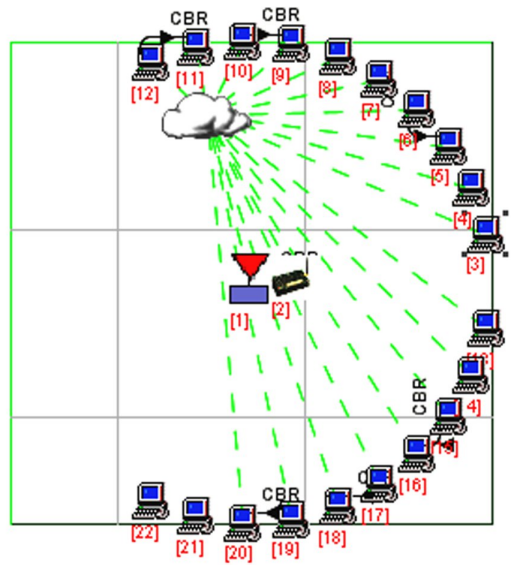

Figure 2.a Coexistence Heterogeneous network scenario for Circular Topology

The figure 2.a shows the scenario of Coexistence Heterogeneous network scenario for Circular Topology developed in Qualnet 4.5 simulator. In this scenario 2 WPAN nodes and 20 WLAN nodes are used. The node 1 is an End device sending packets to the PAN coordinator which is numbered as 2.The WLAN nodes are placed at equal distance about 5 m from the PAN coordinator. From 20 nodes only 5 nodes are assumed as transmitting nodes.

Multiple transmissions in WLAN cause collision which severely affects the performance of WPAN node. The node 1 is RFD (Reduced Functional Device) and node 2 is FFD (Fully Functional Device).The figure 2.b shows the Qualnet scenario for Grid topology with 2 WPAN nodes and 20 WLAN nodes. In grid topology nodes are placed at equal distance from one another .The distance between 2 nodes are fixed as 2m.

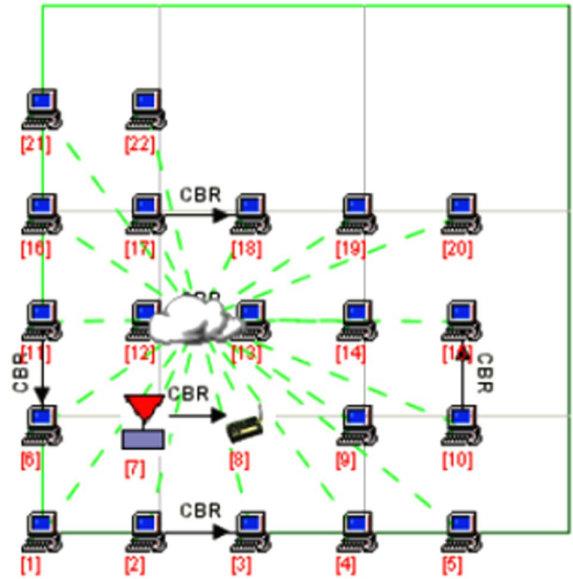

Figure 2.b Coexistence Heterogeneous network scenario for Grid Topology

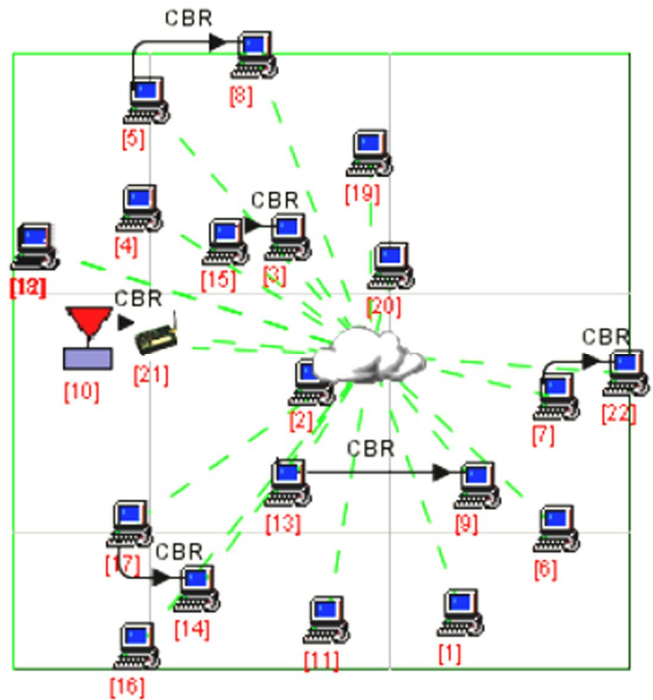

Figure 2.c Coexistence Heterogeneous network scenario for Random Topology

The figure 2.c shows the Qualnet scenario for Random topology with 2 WPAN nodes and 20 WLAN nodes. For this topology seed value is taken as 5.All the nodes are placed randomly except the WPAN nodes. In random topology the distance between the WPAN nodes (node id 10 and 21) are fixed as 1m.

## 4. SIMULATION RESULTS AND DISCUSSION

To evaluate the effectiveness of the proposed scheme in a coexistence heterogeneous wireless network, a simulation study was conducted using Qualnet 4.5 simulator. The simulation configuration and parameters used in this paper is shown in Table 2.

Table 2.Simulation Configuration and Parameters

| Parameter | IEEE 802.11b | IEEE 802.15.4 |
|---|---|---|
| Number of Nodes | 20 | 2 |
| Transmission Power | 20dbm | 3dbm |
| Modulation | CCK | OQPSK |
| MAC Protocol | 802.11 | 802.15.4 |
| Routing Protocol | Bellman ford | AODV |
| No of Packets | 100 | 100 |
| Payload Size | 1500bytes | 105bytes |
| Simulation Time | 100s | |
| Packet Interval | 1s | 0.1 to 1s variable |
| Test bed size | 10m × 10m | |

To study the impact of coexistence on the performance of the 802.15.4 network, measurements were made in a simple experimental environment as shown in Fig. 2(a-c). The effectiveness of the proposed scheme was measured with four different metrics: Data received with errors, Throughput, Average End-End delay and Average jitter. The figure 3(a-d) shows the performance of 802.15.4 network for the four different metrics by varying the packet interval time of 802.15.4 transmission.

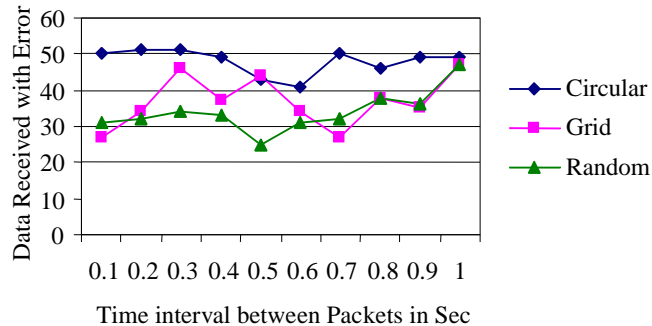

Figure 3.a Data received with errors for Circular, Grid and Random Topology

The error detail of the received data for various topologies is shown in figure 3.a.Among the three topologies the random topology produces minimum error at the packet interval time 0.5 sec. When the packet interval time is equal to the packet interval time of 802.11 network i.e 1sec, all the three topologies produces equal and maximum error. The error is not linearly varying when the packet interval time is varied for all the topologies mentioned.

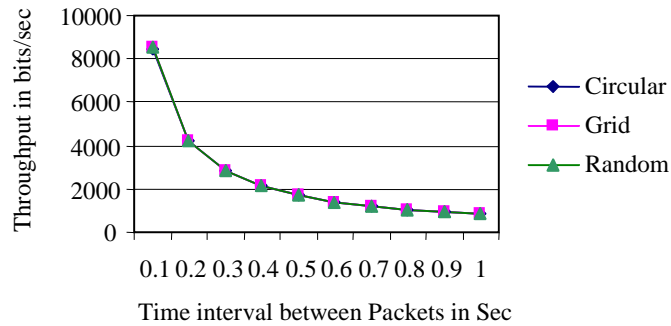

Figure 3.b Throughput for Circular, Grid and Random Topology

The figure 3.b shows the throughput for Circular, Grid and Random Topology by varying the packet interval time from 0.1 sec to 1sec.For all the topologies the throughput value is measured as equal and linearly decreased when the time interval between packet transmissions increases.

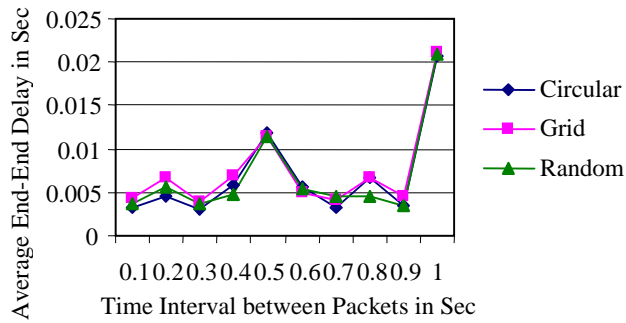

Figure 3.c Average End-End delay for Circular, Grid and Random Topology

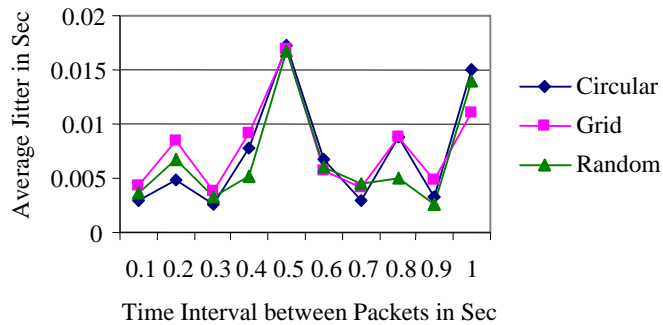

Figure 3.d Average Jitter for Circular, Grid and Random Topology

The average end-end delay and average jitter is shown in figure 3.c and 3.d respectively. The average end-end delay is maximum for all the topologies when the packet interval time is 1sec.The average jitter value is maximum when the time interval between the packets is 0.5sec, at which the random topology produced the minimum error.

The simulation study is extended with the mobility model. The mobility model chosen in this paper is random way point and the speed is fixed as 10m per sec. The figure 4(a-d) shows the performance of 802.15.4 network with the assumption of mobility model.

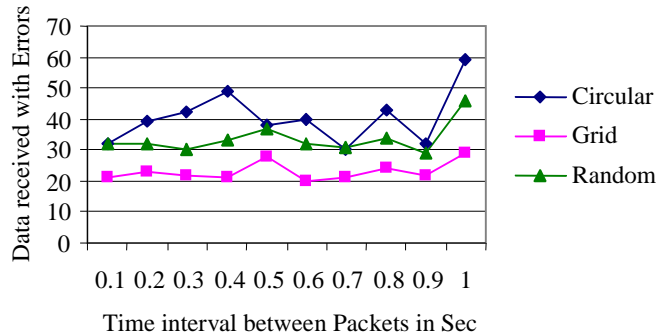

Figure 4.a Data received with errors for Circular, Grid and Random Topology

The figure 4.a shows the error details of the 802.15.4 network for various topologies with the mobility model. Among the topologies circular topology produces more error and the grid topology produces minimum error. The network produces minimum error when compared with the static one for all the topologies. When the packet interval time is equal to the packet interval time of 802.11 network i.e 1sec, all the three topologies produces equal and maximum error.

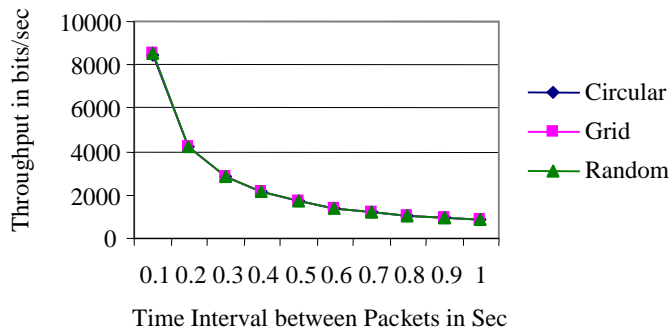

Figure 4.b Throughput for Circular, Grid and Random Topology

The throughput details are shown in figure 4.b.For both static and mobile networks the throughput is equal for all the topologies. The throughput value is measured as equal and linearly decreased when the time interval between packet transmissions increases.

The average end-end delay and average jitter is shown in figure 4.c and 4.d respectively. The average end-end delay is maximum for all the topologies when the packet interval time is 1sec.The average jitter value is maximum when the time interval between the packets is 0.5sec, at which the random topology produced the minimum error. The average end-end delay and average jitter is maximum for grid topology when compared to other topologies. In circular topology the average end-end delay and average jitter is maximum at the inter packet delay 1s.When the same packet interval time for both IEEE 802.11b and IEEE 802.15.4 is fixed the scenario produce more error, maximum average end-end delay and maximum average jitter.

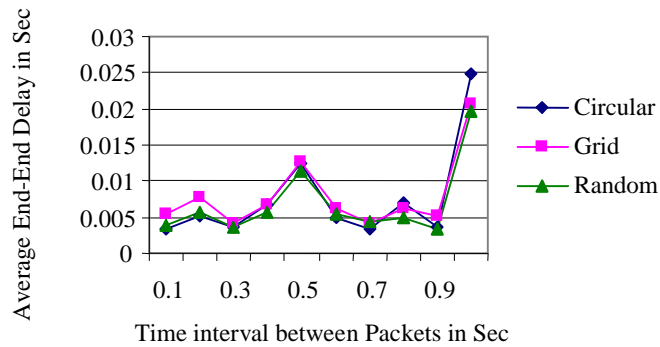

Figure 4.c Average End-End delay for Circular, Grid and Random Topology

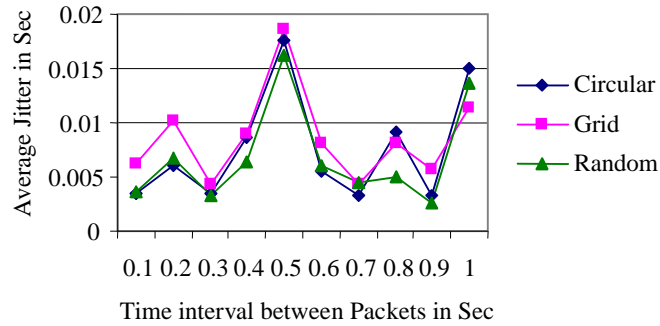

Figure 4.d Average Jitter for Circular, Grid and Random Topology

## 5. CONCLUSIONS

We in this paper present analysis on performance of coexistence heterogeneous networks.In this paper, we propose a new scheme using packet interval time for the coexistence of IEEE 802.15.4 LRWPAN and IEEE 802.11b WLAN.The performance of IEEE 802.15.4 network is analyzed when the nodes are moving randomly. The simulation results show that the proposed scheme is effective in performance improvement for coexistence network of IEEE 802.15.4 for circular, grid and random topologies. In future the analysis can be extended for grid, circular and random topology with the consideration of threading in packet transmission for coexistence heterogeneous networks. Interference mitigation techniques can be incorporated with this scenario for error free transmission.

**Authors**

G.M.Tamilselvan received BE Degree in Electronics and Communication Engineering from Tamilnadu College of Engineering, Coimbatore in 1998 and ME Degree in Process Control and Instrumentation from Faculty of Engineering and Technology, Annamalai University in 2004.From 1998 to 2002 he worked as a faculty in NIIT.He was working as Lecturer in the department of ECE, Erode Sengunthar Engineering College during 2004-2007. Currently he is working as Lecturer in the Department of ECE, Bannariamman Institute of Technology, Sathyamangalam.He is doing part time research in Anna 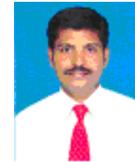 University, Coimbatore.His current research focuses on wireless heterogeneous networks, Interference mitigation, Mesh networks and Adhoc networks. He is member of ISTE and IACSIT, Singapore.
E-mail:tamiltamil@rediffmail.com

Dr.A.Shanmugam received the BE Degree in PSG College of Technology in 1972, Coimbatore and ME Degree from College of Engineering, Guindy, Chennai in 1978 and Doctor of Philosophy in Electrical Engineering from Bharathiar University, Coimbatore in 1994.From 1972–76, he worked as Testing Engineer in Testing and Development Centre, Chennai. He was working as a Lecturer Annamalai University in 1978. He was the Professor and Head of Electronics and Communication Engineering Department at PSG College of Technology, Coimbatore during 1999 to 2004. Authored a book titled 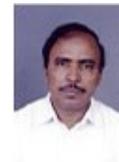 "Computer Communication Networks" which is published by ISTE, New Delhi, 2000.He is currently the Principal, Bannari Amman Institute of Technology, Sathyamangalam.He is on the editorial board of International Journal Artificial Intelligence in Engineering & Technology (ICAIET), University of Malaysia, International Journal on "Systemics, Cybernetics and Informatics (IJSCI)" Pentagram Research Centre, Hyderabad, India. He is member of the IEEE, the IEEE computer society.
E-mail : dras@yahoo.co.in